\documentclass{sig-alternate-2013}

\permission{Copyright is held by the author/owner(s).}
\conferenceinfo{WWW'16 Companion,}{April 11--15, 2016, Montr\'eal, Qu\'ebec, Canada.} 
\copyrightetc{ACM \the\acmcopyr}
\crdata{978-1-4503-4144-8/16/04. \\
Include the http://DOI string/url 
}

\usepackage{hyperref}
\usepackage{times}

\begin{document}

\title{BotOrNot: A System to Evaluate Social Bots}

\numberofauthors{5} 
\author{
\alignauthor
Clayton A. Davis\textsuperscript{1,\S}
\titlenote{Contact: \texttt{claydavi@indiana.edu}\\
{\S}Authors contributed equally.}
\alignauthor
Onur Varol\textsuperscript{1,\S}\\
\and
\alignauthor
Emilio Ferrara\textsuperscript{2}\\
\alignauthor
Alessandro Flammini\textsuperscript{1}\\
\alignauthor
Filippo Menczer\textsuperscript{1}\\
\and       
\affaddr{\textsuperscript{1}Center for Complex Networks and Systems Research, Indiana University, Bloomington, IN (USA)}\\
\affaddr{\textsuperscript{2}Information Sciences Institute, University of Southern California, Marina del Rey, CA (USA)}
}
\maketitle
\begin{abstract}
While most online social  media accounts are controlled by humans, these platforms also host automated agents called social bots or sybil accounts.
Recent literature reported on cases of social bots imitating humans to manipulate discussions, alter the popularity of users, pollute content and spread misinformation, and even perform terrorist propaganda and recruitment actions. 
Here we present \emph{BotOrNot,} a publicly-available service that leverages more than one thousand features to evaluate the extent to which a Twitter account exhibits similarity to the known characteristics of social bots.
Since its release in May 2014, \emph{BotOrNot} has served over one million requests via our website and APIs.
\end{abstract}


\section{Introduction}

A \emph{social bot}, also known as a \textit{sybil account}, is a computer algorithm that automatically produces content and interacts with humans on social media. These agents and their interactions have been observed in online social media for the past few years~\cite{lee2011seven,boshmaf2013design}. Recenly DARPA organized a bot detection challenge to developed techniques for early detection of malicious organized activities~\cite{subrahmanian2016darpa}. 
Some bot accounts are entertaining, helpful, or at least harmless, but nefarious uses for social bots abound, especially when multiple bot accounts are used in a coordinated fashion to perform an orchestrated campaign.
The adoption of social bots has been reported for the purpose of astroturf, that is creating the illusion of artificial grassroots support for political aims~\cite{ratkiewicz2011truthy}.
In another case, a bot campaign created fake ``buzz'' about a tech company: automated stock trading algorithms acted on this chatter, resulting in a spurious 200-fold increase in market price.\footnote{The Curious Case of Cynk, an Abandoned Tech Company Now Worth \$5 Billion --- \url{mashable.com/2014/07/10/cynk}}
An extensive review about social bots and their roles in online social networks is presented in a forthcoming article~\cite{ferrara2014rise}.

In this paper, we present \emph{BotOrNot}, our platform to evaluate whether a Twitter account is controlled by human or machine.
This service is publicly available via the website\footnote{\url{truthy.indiana.edu/botornot}} or via Python or REST APIs.\footnote{\url{github.com/truthy/botornot-python}}$^,$\footnote{\url{truthy.indiana.edu/botornot/rest-api.html}}
\emph{BotOrNot} takes a Twitter screen name, retrieves that account's recent activity, then computes and returns a bot-likelihood score. For website users, this score is accompanied by plots of the various features used for prediction purposes. API tutorials can be found at the pages linked in the footnotes.

\begin{figure}[t!]
    \centering
    \includegraphics[width=0.7\columnwidth]{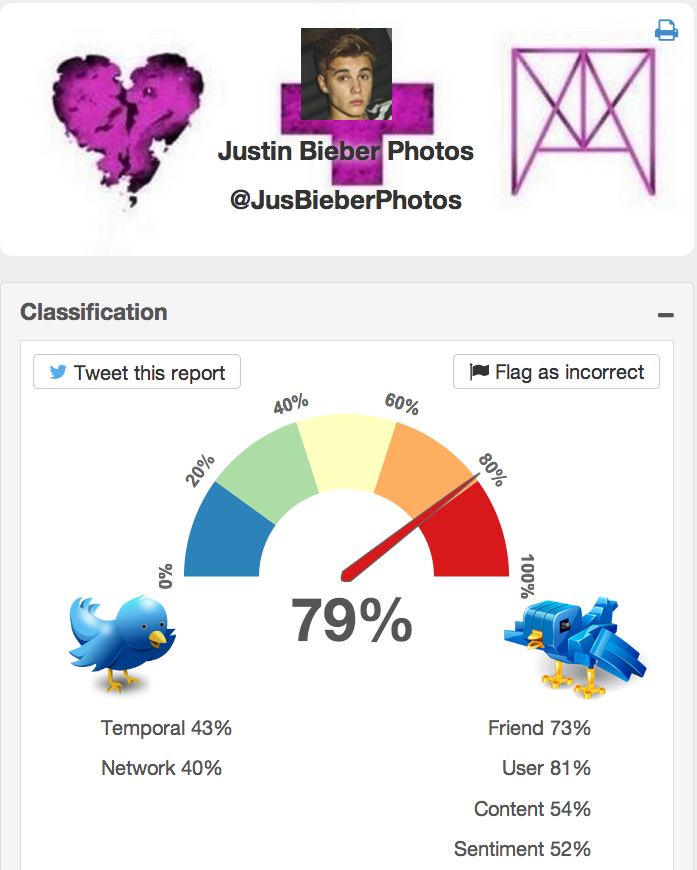}
    \vspace{-1em}
    \caption{\emph{BotOrNot} classification scores interface}
\label{fig:system_design}
\end{figure}

\section{Release Timeline}

We made the \emph{BotOrNot} web service public in May 2014.
Initially our service was only available for users via the website --- there was no public API due to capacity concerns.
With the help of some press coverage, the service was used about 18k times in the first eight months.
As part of a larger effort to address robustness issues causing occasional downtime, we noticed that certain IP addresses were using the service markedly more than others, so we implemented rate limits.
System stability increased as a result of these changes.
The added uptime had the unexpected consequence of increasing overall volume of use.

Analysis of usage by power users revealed that they were using the non-public internal API endpoint for the website.
After a period of serving over 8k requests per day, we decided to explicitly allow programmatic access to \emph{BotOrNot}.
On 11 Dec, 2015, the \textit{@TruthyBotOrNot} Twitter account announced the availability of our public API endpoint with higher rate limits.
In the month since, we have served over 540k requests, bringing the total to over a million queries so far.


\section{System Design}

\subsection{BotOrNot Service}

The use of the \emph{BotOrNot} service starts with a client specifying a Twitter screen name.
The \emph{BotOrNot} website and API use Twitter's REST API\footnote{\url{dev.twitter.com/rest/public}} to obtain the account's recent history, including recent tweets from that account as well as \textit{mentions} of that screen name.
Users are required to have a Twitter account in order for \emph{BotOrNot} to make requests to Twitter's REST API on their behalf. Our API matches Twitter's rate limit of 180 requests per 15 minutes.
Once the requested data is received from Twitter's API, the \emph{BotOrNot} website or API forwards it to the \emph{BotOrNot} server.

The server computes the bot-likelihood score using the classification algorithm described below.
If the request originates from the website, the server  generates data for the plots to be displayed and returns the resulting report with plots (Fig.~\ref{fig:system_design}).
API users receive the classification results in JSON format suitable for post-processing.

\begin{figure}
    \centering
    \includegraphics[width=0.9\columnwidth]{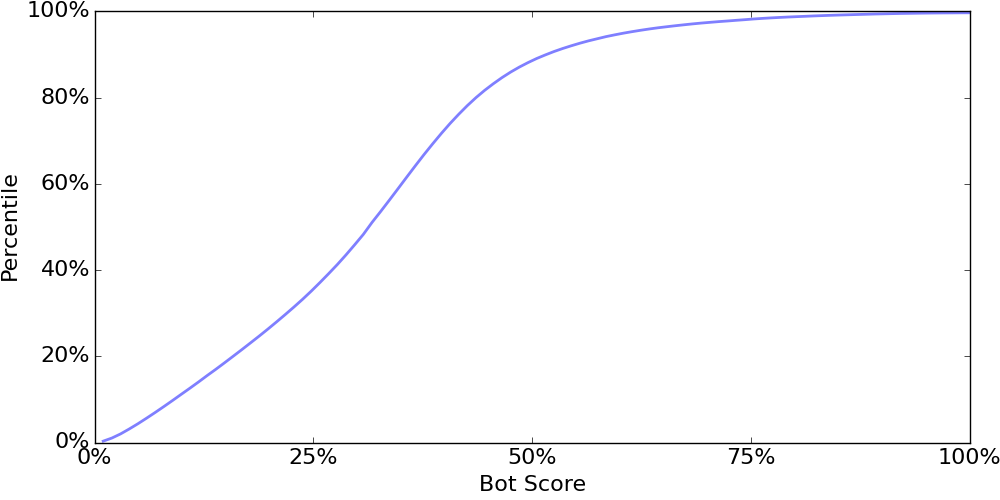}
    \vspace{-1em}
    \caption{Cumulative distribution of bot scores}
\label{fig:bot_score_cdf}
\end{figure}

While \emph{BotOrNot} does not collect data about the users submitting the requests, we do store the computed classification results. As a result, we now have over 900k unique user account classifications. A distribution of bot scores is shown in Fig.~\ref{fig:bot_score_cdf}. We plan to use these collected results to improve the classifier in the future.

\subsection{Classification System}

\emph{BotOrNot}'s classification system generates more than 1,000 features using available meta-data and information extracted from interaction patterns and content. We can group our features into 6 main classes:
\textbf{Network} features capture various dimensions of information diffusion
patterns. We build networks based on retweets, mentions, and hashtag co-occurrence, and extract their statistical features, \textit{e.g.}\ degree distribution, clustering coefficient, and centrality measures.
\textbf{User} features are based on Twitter meta-data related to an account, including language, geographic locations, and account creation time.
\textbf{Friends} features include descriptive statistics relative to an account's social contacts, such as the median, moments, and entropy of the distributions of their number of followers, followees, posts, and so on.
\textbf{Temporal} features capture timing patterns of content generation and consumption, such as tweet rate and inter-tweet time distribution.
\textbf{Content} features  are based on linguistic cues computed through natural language processing, especially part-of-speech tagging.
\textbf{Sentiment} features are built using general-purpose and Twitter specific sentiment analysis algorithms, including happiness, arousal-dominance-valence,
and emoticon scores.

To classify an account as either social bot or human, the model is trained with instances of both classes.
As a proof of concept, we used the list of social bots identified by Caverlee's team~\cite{lee2011seven}.
We used the Twitter Search API to collect up to 200 of their most recent tweets and up to 100 of the most recent tweets mentioning them.  
This procedure yielded a dataset of 15k manually verified social bots and 16k legitimate (human) accounts.
We used this dataset consisting of more than 5.6 millions tweets to train our models and benchmark classification performance.

\emph{BotOrNot}'s classifier uses Random Forest, an ensemble supervised learning method. Extracted features are leveraged to train seven different classifiers: one for each subclass of features and one for the overall score. 
Ten-fold cross-validation yields a performance of 0.95 AUC (Area Under  ROC Curve). Note that such a high accuracy is likely to overestimate current performance, given the age of the training data.

%

\section{Conclusion}

In offering a free social bot evaluation service, we aim to lower the entry barrier for social media researchers, reporters, and enthusiasts.
Ready-made reports on individual users are available via our website, or one can use our API to easily check multiple accounts, up to the rate limit.
While using the API does require some scripting experience, using our service lets users skip the significant step of setting up their own classifiers.

One example application for our service would be a browser plugin adding a context menu option to fetch the \emph{BotOrNot} report for a selected Twitter username.
We welcome such applications from the social media community on top of our public bot classification service.


\textbf{Acknowledgments.} 
This work was supported in part by NSF (grant CCF-1101743), DARPA (grant W911NF-12-1-0037), and the J.S. McDonnell Foundation.

\bibliographystyle{abbrv}
\bibliography{references}

\end{document}